# Phonon Heat Transport and Anisotropic Tuning of Quantum Fluctuations in a Frustrated Honeycomb Magnet

Haoran Fan,[1,2] Yue Chen,[1] Yuchen Gu,[1] Yuan Li,[1,*] and Xi Lin [1,3,4,†]

[1]*International Center for Quantum Materials, School of Physics, Peking University, Beijing 100871, China*
[2]*Beijing Academy of Quantum Information Sciences, Beijing 100193, China*
[3]*Hefei National Laboratory, Hefei, 230088, People's Republic of China*
[4]*Interdisciplinary Institute of Light-Element Quantum Materials and Research Center for Light-Element Advanced Materials, Peking University, Beijing 100871, China*

Honeycomb cobalt oxides containing $3d$ $Co^{2+}$ ions might realize frustrated magnetism and novel quantum phases. Among candidate materials, $Na_3Co_2SbO_6$ stands out for its distorted honeycomb lattice and significant in-plane anisotropy, motivating vector-field tuning inside the honeycomb plane. Here we use thermal transport down to the mK regime to study twin-free crystals of $Na_3Co_2SbO_6$ subject to in-plane vector fields. We find that the thermal conductivity $\kappa$ never exceeds the heat-transport capability of phonons, rendering its suppression primarily due to phonon scattering off magnetic excitations and/or domain boundaries. The system's field-driven quantum criticality manifests itself as an abundance of magnetic fluctuations hindering the heat transport, which further depends on the field direction in an intriguing manner.

Quantum spin liquids (QSLs) are novel states of matter where long-range magnetic order is suppressed by strong quantum fluctuations at zero temperature [1,2]. The exactly solvable spin-1/2 Kitaev honeycomb model provides an ideal platform to study QSLs, and has motivated substantial theoretical and experimental research in recent years [3,4]. Materialization of the Kitaev model has focused on transition metal Mott insulators that contain $5d/4d$ ions, such as $A_2IrO_3$ ($A$ = Na, Li, and Cu) [5,6] and α-$RuCl_3$ [7-14]. The $Ir^{4+}$ and $Ru^{3+}$ ions feature a $d^5$ low-spin configuration $(t_{2g})^5(e_g)^0$ under an octahedral crystal field [15]. In the presence of strong spin-orbit coupling (SOC), the low-energy degrees of freedom can be represented by a spin-orbit entangled $J_{eff} = 1/2$ pseudospin, which exhibits anisotropic Kitaev interactions between the nearest neighbors [16,17]. However, the presence of other types of interactions, including those between further neighbors, as well as the departure from an ideal octahedral crystal field, can often lead to formation of long-range magnetic order [16,18]. While such order might be suppressed by external tuning like strain [19], chemical doping [20], and magnetic fields [10,11,14], achieving a Kitaev QSL ground state has remained a challenge, in part because of stacking and chemical disorders in these materials [8,12,13].

Recently, a new platform has been proposed for realizing Kitaev QSLs with $3d^7$ $Co^{2+}$ ions. Despite the weaker SOC compared to the $4d$ and $5d$ counterparts, the high-spin $(t_{2g})^5(e_g)^2$ configuration also allows for $J_{eff} = 1/2$ physics [21,22], along with mechanisms for suppressing nearest-neighbor Heisenberg and further neighbor interactions [23]. Prominent candidate materials include $Na_2Co_2TeO_6$ [24-29], $BaCo_2(AsO_4)_2$ [30-32], and $Na_3Co_2SbO_6$ [33-36]. Even though they possess magnetic order at low temperatures, it is believed that suppression of the order by in-plane magnetic fields [24,25,32] might still be a promising direction to explore.

$Na_3Co_2SbO_6$ (NCSO) possesses a well-defined monoclinic ($C2/m$) crystal structure [37,38]. At 2 K, magnetometry experiments [33] indicate a relatively small but highly anisotropic in-plane saturation field, ranging from 0.81 T to 1.75 T for fields along the ***b***- and ***a***-axes, respectively. This anisotropy is a clear manifestation of the $C_2$ symmetry, lowered from $C_6$ of an undistorted honeycomb lattice. Meanwhile, neutron diffraction reveals that the magnetic ground state features a double-**q** order [34], instead of the widely explored single-**q** zigzag order [35,39,40] or the triple-**q** order [29] in the sister compound $Na_2Co_2TeO_6$ with $C_6$ symmetry. The structural symmetry is crucial to consider for the purpose of field inducing Kitaev QSLs, because it determines the directions in which an external field may be able to give rise to a gapless QSL state [17,41]. The low symmetry and low saturation field of NCSO thus motivate us to investigate the effects of in-plane magnetic field.

In this work, we present ultra-low temperature thermal transport measurements of twin-free single crystals of NCSO subject to in-plane vector fields. Our results show that the thermal conductivity $\kappa$ is primarily contributed by phonons, which frailly experience scattering off magnetic domain boundaries and low-energy excitations. This makes $\kappa$ a powerful probe of the magnetism. Our key observation is associated with a critical region located between the low-field antiferromagnetic (AFM) states and the high-field polarized ferromagnetic (FM) state, which manifests itself as a "trench" of $\kappa$ minima versus the in-plane fields. Notably, the depth of the trench is anisotropic, which reflects the anisotropic abundance of low-energy magnetic excitations when the system is driven to be least stable by the external field.

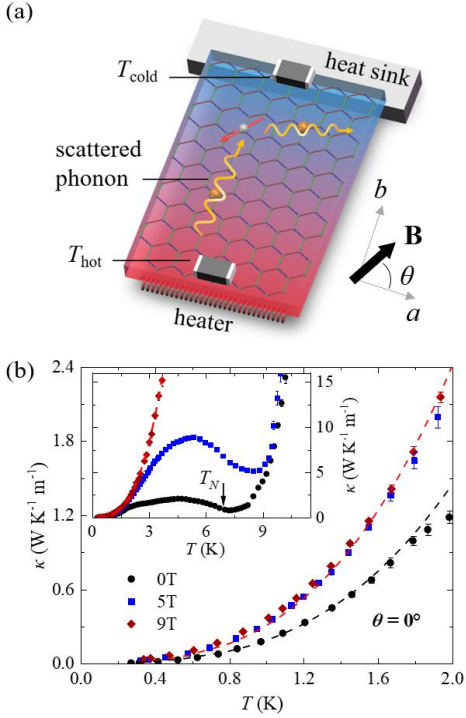

FIG. 1. (a) Schematic of the in-plane rotational thermal transport setup. Yellow wavy lines depict spin-phonon scattering. Inset in the bottom right corner defines the coordinate system for the external magnetic field $B$, where $\theta$ represents the angle of $B$ from the $\boldsymbol{a}$-axis. (b) Field dependence of in-plane thermal conductivity along the $\boldsymbol{a}$-axis from 0.25 K to 2 K. Dash lines represent cubic fits to the data. Inset extends the data to higher temperatures for 0 T, 5 T and 9 T, with the Néel temperature $T_N \approx 6.8$ K marked by an arrow.

We have used previous methods [33] to grow and identify twin-free single crystals of NCSO. Direct current (dc) magnetic susceptibility measurements revealed a well-defined $T_N$ of about 6.8 K. Figure 1(a) illustrates our experimental setup, where the thermal gradient ($\nabla T$) is established along the $\boldsymbol{b}$-axis. The electrical connections are designed to be adiabatic, and the entire setup is mounted on a piezo-rotator (details in Fig. S1 in Ref. [42]), which enables us to apply magnetic fields in all in-plane directions.

In the zero-field magnetic ground state, the system does not have magnon excitations below a gap that is greater than 1 meV [43,44]. Similarly, magnons are expected to be gapped in the high-field polarized state. We can thus expect in both cases to see phonon-dominated heat transport at very low temperatures such that the magnons cannot be excited. This expectation is confirmed by the data in Fig. 1(b), which displays cubic temperature dependences $\kappa(T) = \beta T^3$ characteristic of phonon transport up to 2 K. The saturation of values for fields above 5 T (see Fig. S2(b) for $\boldsymbol{b}$-axis fields in Ref. [42]), with $\beta = 0.31$ W/(K·m), indicates that the magnon gap is large enough to prevent thermal excitation up to 2 K. The inset of Fig. 1(b) shows that a greater field of 9 T lets the $T^3$ dependence persist up to higher temperatures. We further compare this result to the non-magnetic isostructural material Li$_3$Zn$_2$SbO$_6$ [45], where the phonon heat capacity of about $c_{\mathrm{ph}} = 2.56 \times 10^{-4} T^3$ J/(mol·K) corresponds to a sound velocity of about 1468 m/s. In the ballistic limit [46], the phonon mean free path is determined by the cross section area ($A$) of the sample, $l_{\mathrm{ph}} = 2\sqrt{A/\pi} \approx 0.56$ mm in our case. Thus, we estimate the phononic thermal conductivity to be $\kappa_{\mathrm{ph}} = \beta_0 T^3 \approx 0.89 T^3$ W/(K·m), which is not far above our observed values at high fields, lending support to our phonon hypothesis.

We therefore consider $\kappa_{\mathrm{ph}}$ measured at the field of 9 T an important reference, in the light of the spin-phonon scattering mechanism described in [11,28]. A closer scrutiny of the data in the inset of Fig. 1(b) (and additional data in Fig. S2 in Ref. [42]) further supports this understanding: thermally activated magnetic excitations suppress $\kappa$ in the intermediate temperature range for the zero-field and 5 T data. Moreover, $\kappa(T)$ increases steeply above the zero-field Néel temperature $T_N$, where the magnetic excitations are expected to play a less significant role. Altogether, these observations strongly suggest that the heat transport is primarily contributed by phonon transport subject to the hindrance of spin-phonon scattering.

We next turn to isothermal traces of $\kappa(B)$. In Fig. 2(a), the trace for $B \parallel \boldsymbol{a}$ obtained at 285 mK saturates above 5 T, which corresponds to the intrinsic phonon contribution. At lower fields, $\kappa(B)$ undergoes a variety of suppressions, which can be divided into three regions for detailed inspection.

Region I is the long-range ordered state at low fields [35] with gapped magnons [36,43,44]. Even though the temperature dependence is proportional to $T^3$ [Fig. 1(b)], $\kappa$ here is far below the intrinsic phononic $\kappa_{\mathrm{ph}}$ at high fields. The suppression resembles behaviors observed in α-RuCl$_3$ [47], and we attribute it to the scattering of phonons by magnetic domain boundaries, which will be discussed later.

Region II marks a local minimum of $\kappa(B)$ at the transition field $B_{c1}$ of 1.4 T between two AFM orders, where propagating wave vectors switch from $q_{AFM1/2} = \pm \boldsymbol{a}^*/2 \pm \boldsymbol{b}^*/2$ to $q_{AFM1/3} = \pm \boldsymbol{a}^*/3 \pm \boldsymbol{b}^*/3 \pm \boldsymbol{c}^*/3$ [33]. The location of this transition depends on the field direction, and is at 0.58 T along the $\boldsymbol{b}$-axis [inset of Fig. 2(a)]. Since the transition is known to be first-order [33], the suppression of $\kappa$ can be attributed to phonon scattering on the first-order phase boundaries and/or energy dissipation associated with the hysteretic phase transition [48,49]. While the observation may be related resembles thermal transport behaviors near critical regions in related systems [10,28], and concurs with the presence of strong quantum fluctuations in NCSO [36,50]. Therefore, our data suggest that the magnetic quantum fluctuations near $B_{c2}$ do not directly contribute to the thermal conductivity. Instead, their primary role is to suppress phonon heat transport through spin-phonon scattering.

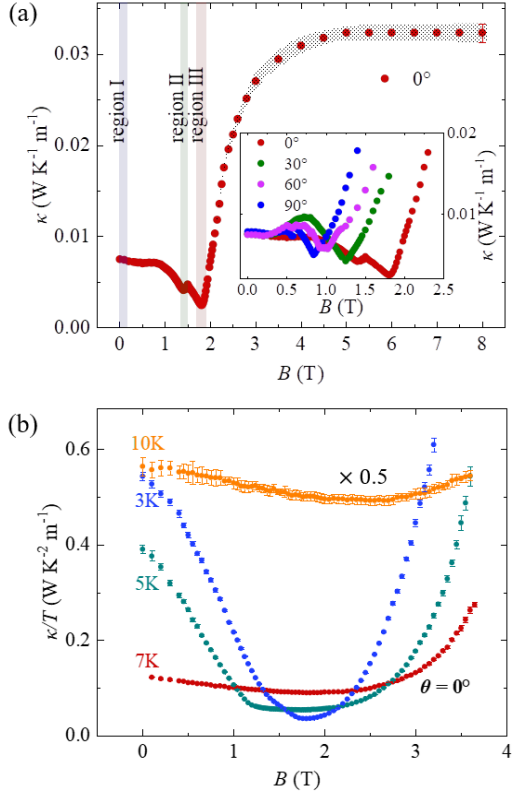

As the temperature increases, the above detailed structures in $\kappa(B)$ become less pronounced (Fig. 2(b), see Fig. S3 for $b$-axis data in Ref. [42]), which indicates that thermal fluctuations overcome the effects produced by the domain and phase boundaries. In contrast, the minimum around $B_{c2} = 1.8$ T persists up to at least 5 K, which highlights the prominence of the critical fluctuations. Intriguingly, the thermal conductivity value in the critical region, denoted as $\kappa(B_{c2})$, displays a non-monotonic dependence on the field orientation [inset of Fig. 2(a)]. Specifically, $\kappa(B_{c2})|_{\theta=90°}$ is nearly twice as large as $\kappa(B_{c2})|_{\theta=0°}$, while the intermediate $\kappa(B_{c2})|_{\theta=60°}$ is even larger than $\kappa(B_{c2})|_{\theta=90°}$. This indicates that the strength of quantum critical spin-phonon scattering in the zero-temperature limit depends on the field orientation in a non-trivial fashion.

In Fig. 3(a), we present the results of our detailed measurements of the full vector-field dependence of $\kappa$ at 285 mK. As expected, a uniform phononic $\kappa_{ph}$ is observed above 5 T, where the difference between different field directions is less than 5%. Our main result is an intriguing "trench" observed below 2 T, which marks a manifold of maximal spin-phonon scattering caused by critical magnetic fluctuations. A zoom-in view below 2 T is shown in Fig. 3(b). In the region near zero field, we notice that $\kappa(0°)$ and $\kappa(90°)$ are considerably smaller than $\kappa(\theta+n\cdot90°)$, where $n = 0, 1, 2, 3$ and $\theta$ ranges from 20° to 70°. This is in part related to the domain-boundary scattering discussed above for the suppression of $\kappa$ in Region I. In fact, we believe that the higher values of $\kappa(\theta+n\cdot90°)$ are related to the field-scanning sequence employed in our measurements at each angle, where the to heat capacity results in Fig. S4(e)-(f) (Ref. [42]), a detailed analysis of this falls beyond our present scope.

FIG. 2. (a) The main panel displays the thermal conductivity $\kappa$ at 285 mK for fields along the $a$-axis from 0 T to 8 T. Regions labelled as I, II, and III correspond to the zero-field state, a first-order phase transition, and a quantum critical regime, respectively. The shadow above 2 T indicates measurement uncertainty arising from the magnetic field dependence of the RuO$_2$ thermometer. Inset displays low-field data at $\theta = 0°$, 30°, 60° and 90°. (b) Magnetic field dependence at four selected temperatures. Data at 10 K are scaled down by 50% for clarity.

Region III is situated between the AFM and the polarized states and features the strongest phonon scattering, resulting in a global minimum at $\kappa(B_{c2})$. It measurements are performed up to 8 T before turning to the next field direction. This appears to introduce a training memory effect and helps increase $\kappa$ measured near zero field, and the increase is more significant for $\kappa(\theta + n \cdot 90°)$ than for $\kappa(0°)$ and $\kappa(90°)$ (see Fig. S4 and Supplementary Note 4 for detail in Ref. [42]). The results suggest that the field training process helps eliminate some of the AFM domain boundaries in the zero-field state, and thus mitigates the scattering of phonons. It is understandable that the lower-symmetry field directions are more effective with the domain alignment.

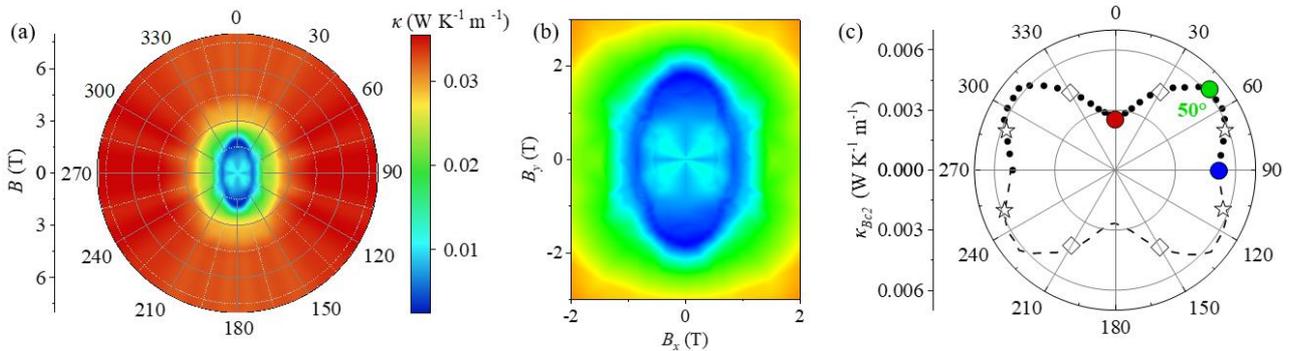

FIG. 3. (a) Vector-field dependence of $\kappa$ at 285 mK presented by false colors in the polar coordinate, ranging from 0 T to 8 T (data in the range of 180° - 360° are symmetrically derived from 0° - 180° raw data, the contour map is plotted via Thin-Plate-Spine interpolation). (b) Zoom-in view of $\kappa$ with in-plane field ranging from 0 T to 2.5 T. (c) Thermal conductivity $\kappa_{B_{c2}}$ at the critical point $B_{c2}$ at 285 mK exhibits angular dependence. Actual measurements have been carried out at angles denoted by circles, diamonds and stars, whereas the dashed lines are a guide for the eye.

Greater caution should be taken when interpreting the minima of $\kappa$ near $B_{c2}$, where no field hysteresis is observed (Fig. S4(a)-(d) in Ref. [42]) and the spin-phonon scattering mainly arises from magnetic critical fluctuations. The significance of these fluctuations is reflected by the angular dependence $\kappa(\theta)|_{B=Bc2}$ presented in Fig. 3(c), where a broad maximum is observed at about 50°. Importantly, the non-monotonic angular dependence has no counterpart in the magnetometry results reported previously [33]. The hump at $\kappa(50°)|_{B=Bc2}$ suggests that the amount of magnetic quantum fluctuations in this field direction is about 20% less than for $B\|\mathbf{b}$, which is a further 50% less than for $B\|\mathbf{a}$.

To further understand the anisotropy of the critical fluctuations at $B_{c2}$, we present in Fig. 4 the anisotropy as a function of temperature, along with the zero-field data as a reference. It is seen that while $\kappa(0°)|_{B=Bc2}$ is the smallest at all temperatures, $\kappa(50°)|_{B=Bc2}$ exceeds $\kappa(90°)|_{B=Bc2}$ below 2 K. As the temperature approaches the zero-field $T_N$, all three orientations converge to the zero-field curve and the anisotropy disappears, which indicates that the high-temperature paramagnetic excitations are much less sensitive to the external field than the excitations at lower temperatures. Therefore, our data unambiguously show that the competition between the AFM and FM correlations [33] is at the origin of the quantum critical fluctuations near $B_{c2}$, such that the critical regime essentially terminates above the zero-field $T_N$. This important nature of the quantum criticality has not been revealed by resonance probes which work at much lower energy [36,50] than the heat-carrying phonons.

The temperature of 2 K appears to be an important turning point for the anisotropy at $B_{c2}$: not only does it mark the crossing between $\kappa(90°)$ and $\kappa(50°)$ (Fig. 4, main panel), but it also marks the maximal anisotropy between $\kappa(90°)$ and $\kappa(0°)$ (Fig. 4, inset). The exact origin of this anisotropy and its evolution with temperature is unclear at this point. Nevertheless, we emphasize that our detection of it via the spin-phonon scattering (and integrating their influence on the thermal transport) likely constitutes an unbiased sampling of all low-energy magnetic fluctuations, and may therefore be a faithful representation of the genuine magnitude of the critical fluctuations. Our results strongly suggest that the strength of the quantum criticality also exhibits pronounced anisotropy in the zero-temperature limit. As a possible interpretation, a variation in the strength of the quantum criticality may be introduced by a weakly first-order nature of the transition at $B_{c2}$ [50], such that the residual signature of the criticality may depend on the anisotropic tri-critical temperature associated with the transition [51]. For a detailed discussion and additional data related to this, see Fig. S5 and Supplementary Note 5 in Ref. [42].

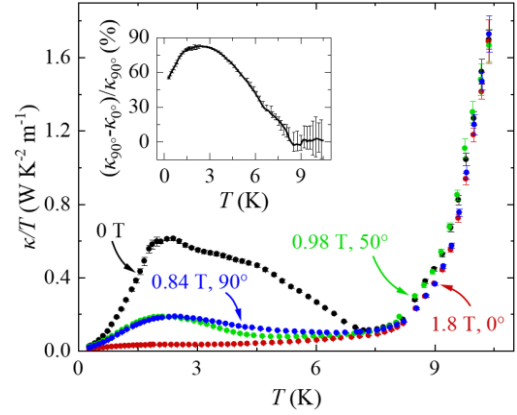

FIG. 4. Temperature dependence of $\kappa/T$ at 0 T and three representative angles under their corresponding critical field $B_{c2}$. The inset displays the normalized thermal conductivity difference between the two principal axes. Error bars reflect the uncertainty attributed to the magnetic field influence of the thermometers.

On a final note, our results highlight NCSO as a rare case of honeycomb quantum magnet which exhibits pronounced in-plane anisotropy in the most interesting quantum critical regime. Such anisotropy, while awaiting thorough theoretical modeling and understanding, has been demonstrated to be absent in other well-studied cobalt oxides [24,28,31]. We note that a recent discovery of a monoclinic polymorph of $Na_2Co_2TeO_6$ [52], which has the same space group $C2/m$ as NCSO, may add a second system to this family that can be effectively tuned towards quantum criticality or even QSLs by in-plane vector fields. Our work therefore illuminates a promising tuning strategy, namely, by using phonon heat transport as an indicator and aiming to minimize the thermal conductivity at ultra-low temperature, perhaps over a more extended range in the field magnitude (if a QSL phase can indeed be realized).

In summary, we have presented a first high-resolution in-plane-rotation thermal transport study of $Na_3Co_2SbO_6$ twin-free crystals. The thermal conductivity is governed by phonons highly sensitive to spin-phonon scattering, which we have utilized to investigate the system's quantum critical behavior at ultra-low temperatures. Our data reveal an intriguing in-plane anisotropy of the strength of the quantum criticality, and will motivate further theoretical and experimental studies of systems that offer the opportunity to be sensitively tuned by in-plane vector fields.

The authors are grateful to Xiaochen Hong, Christian Hess, Xiaoqun Wang, Zheng Liu, Wei Li, Yisheng Chai and Mingquan He for stimulating discussions. This work is supported by the National Key Research and Development Program of China (Grant No. 2021YFA1401901) and the NSF of China (Grant No. 12141001, No. 12061131004, and No. 11921005).

*yuan.li@pku.edu.cn

†xilin@pku.edu.cn

# Supplementary Material for "Phonon Heat Transport and Anisotropic Tuning of Quantum Fluctuations in a Frustrated Honeycomb Magnet"

Haoran Fan,[1,2] Yue Chen,[1] Yuchen Gu,[1] Yuan Li,[1,*] and Xi Lin [1,3,4,†]

[1]*International Center for Quantum Materials, School of Physics, Peking University, Beijing 100871, China*
[2]*Beijing Academy of Quantum Information Sciences, Beijing 100193, China*
[3]*Hefei National Laboratory Hefei, 230088, People's Republic of China*
[4]*Interdisciplinary Institute of Light-Element Quantum Materials and Research Center for Light-Element Advanced Materials, Peking University, Beijing 100871, China*

## 1. IN-PLANE ROTATION THERMAL TRANSPORT CELL AND MEASUREMENT TECHNIQUS

The thermal conductivity of the single crystals $Na_3Co_2SbO_6$ (2.5×1.25×0.2 mm) is measured by our homemade in-situ rotatable thermal transport setup as shown in Fig. S1. The configuration diagram of the setup refers to Fig. S1(a), where the sample is fixed by an insulated fishing line and a strong thermal linked heat sink. A heater and two $RuO_2$ chip thermometers are glued on the sample directly by diluted Ge-Varnish to ensure a good thermal link, and the heat $\dot{q}$ is flowing along the ***b***-axis of 90°. The measuring wires are connected by copper and bare superconductor wires, where the superconductivity can help to insulate the heater from leakage. Figure S1(b) is the corresponding photo, and Fig. S1(c) shows the whole sample holder is mounted on the piezo rotator, which can achieve 360° in-plane rotation. The angle between the *B*-field and the direction of heat is measured by a Hall element.

Figure S1(d) shows the performance of the setup at isothermal $\kappa(B)$ measurement. The time between two dotted lines represents one measurement period, which includes the initial sweeping field of the yellow shadow region, then the balance time, and finally the read data window of the grey shadow region. After each step increase, $T_{Hot}$ and $T_{Cold}$ on the sample reveal dramatic heating transient via the complicated magneto-caloric effect from the sample, while $T_{Base}$ still maintains a nearly constant, proving the adiabaticity of the setup. To guarantee the homogeneity, the thermal gradient $\nabla T$ is set to not exceed 1% of the average sample temperature.

The $RuO_2$ chips are measured by standard lock-in techniques at 17.593 or 19.975 Hz with 10 nA, and they are calibrated *in situ* against another $RuO_2$ thermometer with reliable thermometry. The heater is excited by a constant-current source with less than 10 uW power below 10K. The measurements are carried out in an integrated Cryofree® superconducting magnet system (TeslatronPT, Oxford instrument) and a ³He sample-in-vacuum insert system (HelioxVT, Oxford instrument, 250 mK base temperature). The key observations are highly reproducible in separate cool-downs.

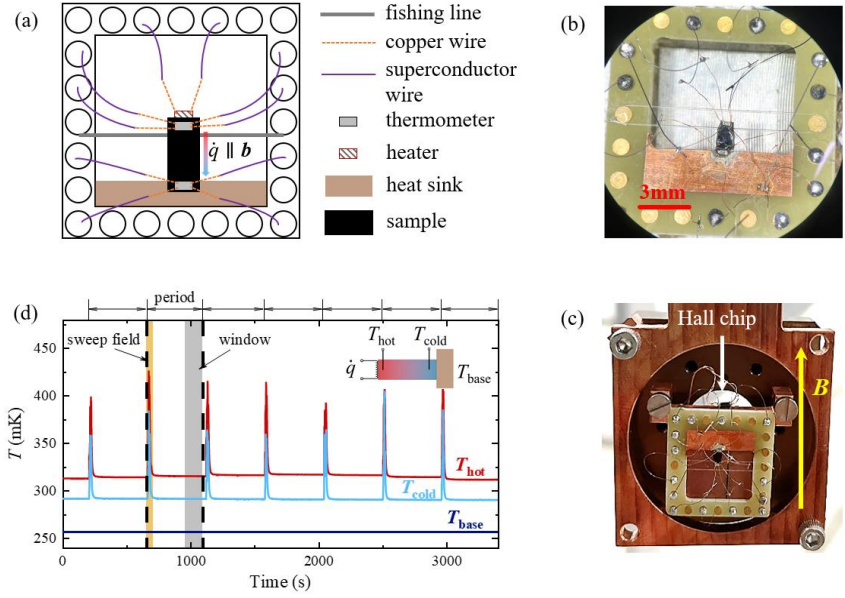

FIG. S1. (a) Detailed schematic of the homemade thermal transport setup. (b) A photo taken by a microscope shows the sample mounted in this measurement unit. (c) A Photo of the unit installed on the piezo rotator and Hall element to determine angle, with the direction of external field. (d) An isothermal $\kappa(B)$ measurement example is taken at 0° between 1.72 T to 1.86 T.

## 2. ADDITIONAL FIELD DEPENDEND DATA OF $\kappa(T)$

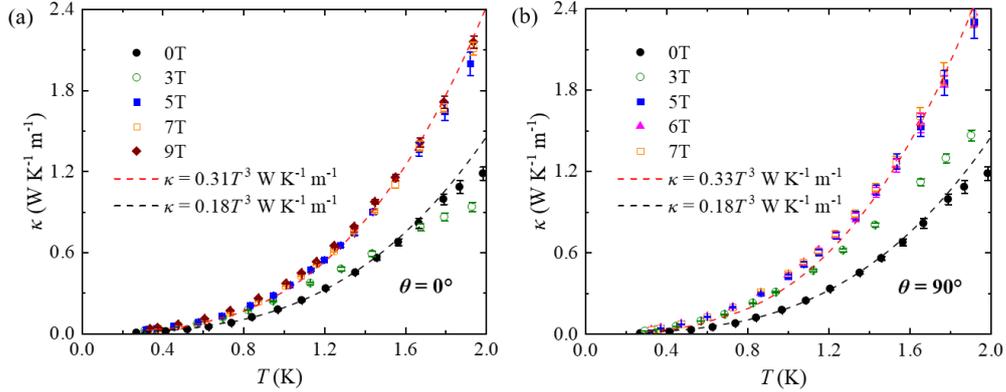

FIG. S2. Field dependence of in-plane thermal conductivity $\kappa(T)$, ranging from 0.25 K to 2 K along ***a***-axis (a) and ***b***-axis (b). The black and red dashes represent the cubic fits of 0 T and high field $\kappa(T)$ (9 T for the ***a***-axis and 7 T for the ***b***-axis).

## 3. ADDITIONAL DATA OF MAGNETOTHERMAL CONDUCTIVITY

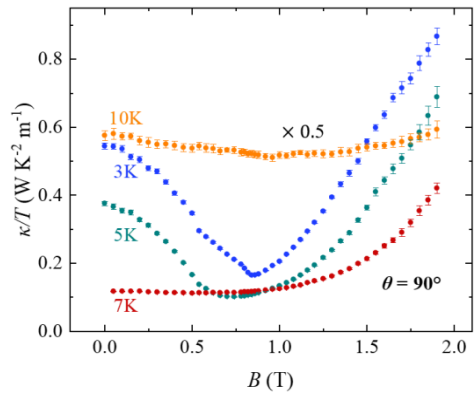

FIG. S3. Magnetic field dependence at four selected temperatures along the ***b***-axis. Data at 10 K are scaled down by 50% for clarity.

## 4. HYSTERETIC BEHAVIOR AT LOW FIELDs

A detailed magnetothermal conductivity $\kappa(B)$ and heat capacity $c(B)$ are conducted to characterize the hysteresis behavior. The $\kappa(B)$ shows distinct hysteresis below $B_{c1}$ along four orientations [Fig. S4 (a)-(d)]. A similar scenario has been discussed in $Na_2Co_2TeO_6$, where the impact of possible magnetic domain structures equally distributed or unequal density of state are speculated in the low fields [24,28]. However, unlike the hysteresis only observed along *armchair* orientation in $Na_2Co_2TeO_6$, hysteresis in NCSO appears along all orientations [Fig. 3(b)], extending to zero fields. An AC calorimetry [53] is used to check the source [Fig. S4 (e)-(f)]. The $c(B)$ occurs a "jump" at $B_{c1}$ during field ramp-up in *a* (1.42 T) and *b* (0.58 T) axes, and the same is observed during the ramp-down process along *b*-axis (0.42 T) but faint along *a*-axis (1.17 T), consistent with the first-order phase transition observed in $\kappa(B)$ and magnetometry [33]. On the contrary, such hysteresis disappears once away from $B_{c1}$, revealing the same density of state on the side of $B_{c1}$ in different processes, and is attributed to some thermodynamically trivial stems that scatter the intrinsic phonons.

We naturally deduce phonons are suppressed by the more conventional mechanism of domain boundaries in long-range order. In this line, a zero-field cooling (ZFC) forms random distributed magnetic domains in the 2D honeycomb lattice, whose size is expected to exceed phonon wavelength. The external field will reverse some of these domains into an unequaled distribution, resulting in fewer domain boundaries, thus increasing low-field $\kappa$.

Following this line, we can estimate the domain boundary's size by fitting with the 0 T data (Fig. 1(b) and Fig. S2). 0 T thermal conductivity also satisfies a cubic temperature dependence below 1.8 K with $\kappa(T)/T = \gamma T^2$ [$\gamma = 0.18$ W/(K·m)], implying the intrinsic phonon even under the boundary scattering. Therefore, we get $l_{\text{domain}} = \gamma/\beta \times l_{\text{ph}} \approx 0.32$ mm.

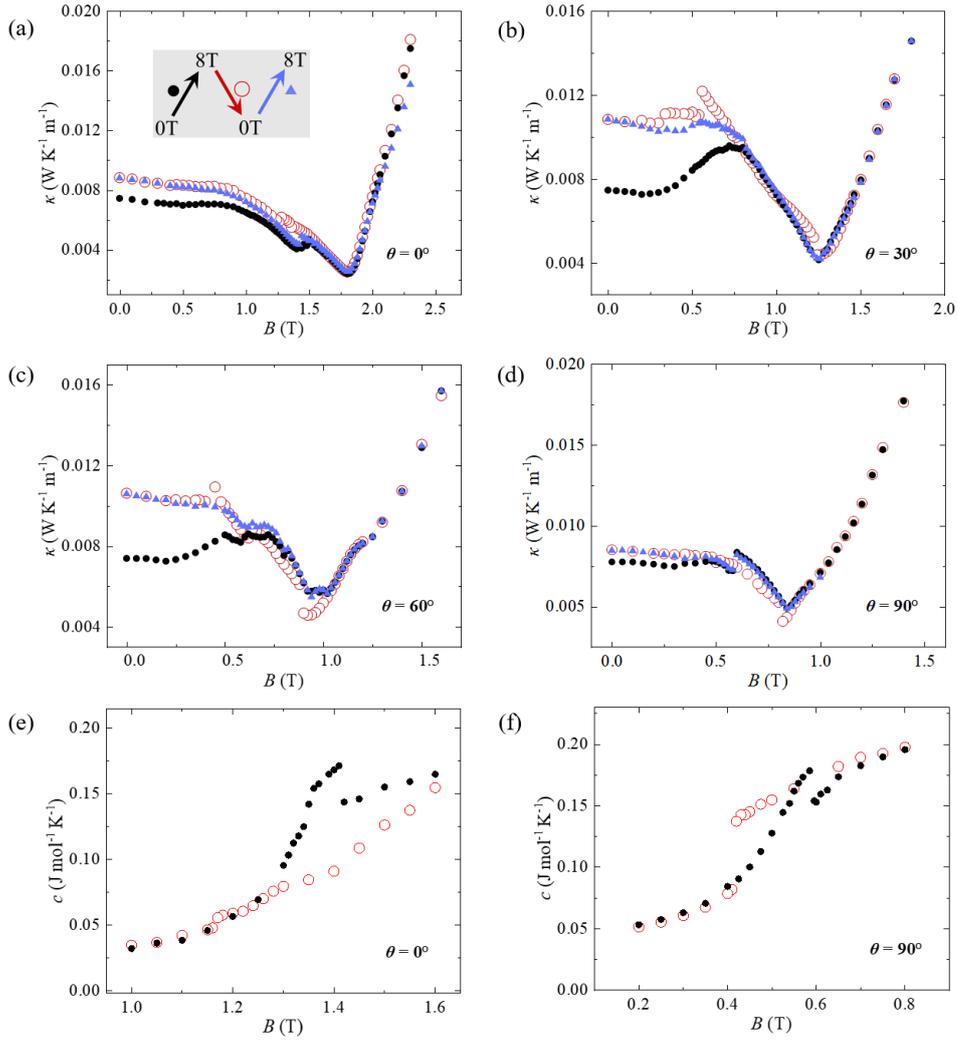

FIG. S4. At 285 mK, the hysteresis behavior of the isothermal $\kappa(B)$ at four orientations of 0°, 30°, 60° and 90° (a-d). Taking 0° (a) as an example: the ramping sequence is shown in the insert which starts from an initial ZFC $\kappa(0\text{ T})$ to $\kappa(8\text{ T})$ (black dots), then ramps down (red open circles), finally ramp up (purple triangles). The whole features as a hysteresis below $B_{c1}$, and is apparent for 30°, 60° (b-c). The data in Fig. 3(a)-(b) is adopted during the final ramp up (purple triangles). Heat capacity $c(B)$ along 0° and 90° around $B_{c1}$ (e-f).

## 5. DIFFERENT CRITICAL BEHAVIORS ALONG TWO PRINCIPAL AXES

The 0 T thermal conductivity shows minimum consistent with the classic paramagnon fluctuations at Néel temperature [Fig. 1(b)]. An intermediate field is predicted to lower such thermodynamics features [36]. We measure a series of magnetothermal conductivity below 6.8 K to investigate the evolution of the "trisecting point" of FM, AFM, and paramagnetic phases. To clarify the $B_{c2}$ dip's feature, every $\kappa(B)$ is scaled to its corresponding critical thermal conductivity $\kappa(B_{c2})$ (extracted by $\partial \kappa(B)/\partial B$).

*a*-axis: Below $B_{c1}$, $\kappa(B_{c2})/\kappa(B)$ decreases with temperature [Fig. S5(a)], indicating less scattering between phonons and domain boundaries due to more activated phonons. Around $B_{c2}$, $\kappa(B_{c2})/\kappa(B)$ shows a close uniform "peak" below 1.35 K, which is attributed to a single scattering origin of the critical fluctuations between FM with AFM. However, such a "peak" broadens above 1.35 K. Combining with the phase diagram [36], we soon notice the phase separation of AFM and FM once the temperature rises. Therefore, a higher temperature will give a wider magnetic fluctuations region, resulting in a more broadened dip near critical field 1.8 T [Fig. 2(b)].

*b*-axis: the suppression below $B_{c1}$ follows the same mechanism as the *a*-axis. However, the "peak" of $\kappa(B_{c2})/\kappa(B)$ shows uniform behavior below 4.4 K [Fig. S5(b)], much higher than *a*-axis, which means the "trisecting point" is estimated to occur at 0.84 T ~4.4 K.

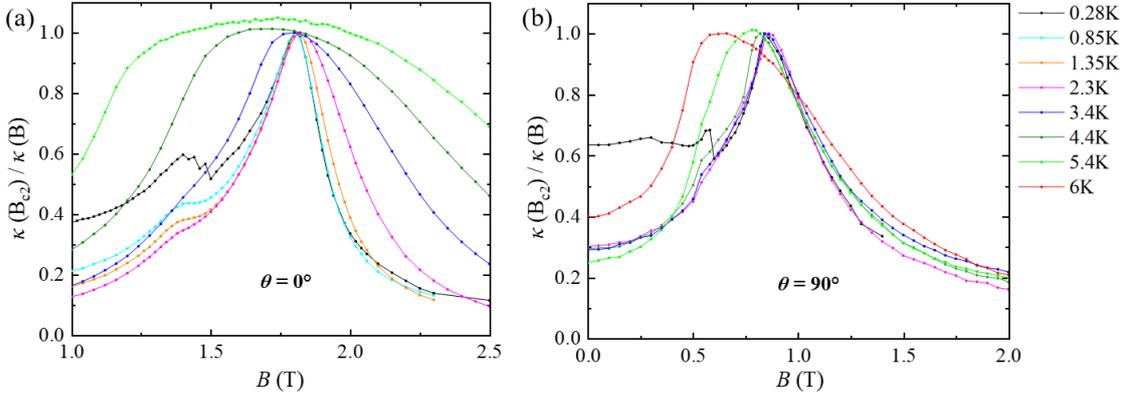

FIG. S5. The magnetothermal conductivity scaled to critical field $B_{c2}$ along *a*-axis (a) and *b*-axis (b).